\documentclass{article}
\usepackage[utf8]{inputenc}
\usepackage[backend=biber, style=ieee, sorting=nty]{biblatex}

\usepackage{amsmath}
\usepackage{amssymb}
\usepackage{graphicx}
\usepackage{fancyhdr}
\title{A Closer Look at the Tropical Cryptography}
\date{}
\author{Steve Isaac and Delaram Kahrobaei}

\fancypagestyle{affiliation}{
\fancyhf{}

\cfoot{Department of Computer Science, University of York, Heslington, York, YO10 5DD, United Kingdom}
}
\begin{document}

\maketitle
\thispagestyle{affiliation}
\begin{abstract}
    We examine two public key exchange protocols proposed recently which use tropical algebra. We introduce a fast attack on the first protocol, and we show that the second protocol cannot be implemented as described.
\end{abstract}

\section{Introduction}
In this paper we analyse the two novel key exchange protocols proposed in \cite{TropCrypt2}, which are both based on tropical matrix algebra. These protocols utilise the semidirect products of semigroups \cite{SemidirectProductofSemigroups}, in an attempt to avoid exploitable patterns, which were exhibited by previous protocols based on tropical matrix algebra \cite{Kotov2015AnalysisOA}. It has already been shown that a private parameter of these novel protocols can be recovered in about 15 minutes using a binary search \cite{RudyArxiv}. We will describe a significantly faster method of attack on the first protocol, which recovers the same private parameter. We will also show that the second protocol cannot be implemented as described, due to its reliance on the associativity of an operation that is not associative.
\section{Preliminaries}
\subsection{Tropical Matrix Algebra}
The tropical algebra, $\overline{\mathbb{R}}$, is given by equipping the extended set of real numbers, \(\mathbb{R} \cup \{\infty\}\), with the addition operation, $\oplus$, and the multiplication operation, $\otimes$, defined by:
\begin{equation}
    x\oplus y=\min(x, y)
\end{equation}
\begin{equation}
    x\otimes y=x+y
\end{equation}
$\overline{\mathbb{R}}$ satisfies all the axioms of a semiring. On top of satisfying the semiring axioms, $\overline{\mathbb{R}}$ is idempotent under addition (and therefore an idempotent semiring) and commutative under multiplication.

A tropical matrix algebra, $\overline{T}$ can be defined by equipping the set of square matrices of order $\omega$, that contains elements from $\overline{\mathbb{R}}$, with the addition operation, $\oplus$, and the multiplication operation, $\otimes$. The elements of the matrix \(X\), produced by the matrix addition \(Y\oplus Z\), are defined by:
\begin{equation}
X_{ij}=Y_{ij}\oplus Z_{ij}
\end{equation}
The elements of the matrix \(X\), produced by the matrix multiplication \(Y\otimes Z\), are defined by:
\begin{equation}
X_{ij}=\bigoplus_{k=1}^\omega Y_{ik}\otimes Z_{kj}
\end{equation}

Like $\overline{\mathbb{R}}$, $\overline{T}$ is an idempotent semiring, however $\overline{T}$ is not commutative under multiplication.

Note that $\overline{\mathbb{R}}$ is also known as min-plus algebra. For further reading on tropical algebra see \cite{Baccelli1994SynchronizationAL} and \cite{Max-linearSystems}.

\subsection{Almost Linear Periodicity}
\citeauthor{MatrixPowers} define a sequence of matrices, $H^n, n \in \mathbb{N}$, as almost linear periodic \cite{MatrixPowers} if there exists a period $\rho$, a linear factor $\xi$, and some defect $d$, such that for all $n>d$ and all indices $i, j$ the following equation holds:
\begin{equation}
    H^{n+\rho}_{ij} =\xi +H^{n}_{ij}
\end{equation}
This property is exhibited by the first protocol, and exploited to create our attack.

\section{Overview of the first protocol}
N\kern-0.3em{B} Throughout this paper we will use $M_n$ and $H_n$ to represent the results of $(M,H)^n$: that is to say \((M,H)^n=(M_n,H_n)\).\\\\
The protocol is based on a semigroup given by equipping the set of pairs, containing entries from $\overline{T}$, with the operation defined by:
\begin{equation}
(X,G)(Y,H)=((X\circ H)\oplus Y, G\circ H)
\end{equation}
where $X\circ H$ is defined by:
\begin{equation}
X\circ H = X\oplus H\oplus (X\otimes H)
\end{equation}

\begin{enumerate}
    \item Alice and Bob agree upon public matrices $M$ and $H$, with elements in $\mathbb{Z}$.
    \item Alice generates a private positive integer $a$, while Bob generates a private positive integer $b$.
    \item Alice calculates \((M,H)^a=(M_a,H_a)\), and shares $M_a$ with Bob, keeping $H_a$ private.
    \item Bob calculates \((M,H)^b=(M_b,H_b)\), and shares $M_b$ with Alice, keeping $H_b$ private.
    \item Alice calculates \(K=(M_b\circ H_a)\oplus M_a\) to get the secret key.
    \item Bob calculates \(K=(M_a\circ H_b)\oplus M_b\) to get the same key as Alice.
\end{enumerate}

When implementing the protocol \citeauthor{TropCrypt2} state that $(M,H)^a$ should be calculated using the square-and-multiply method. The associativity of the semigroup operation makes this possible.
\section{Attack on the first protocol}
The sequence of matrix powers over $\overline{T}$ is shown to be almost linear periodic in \cite{Baccelli1994SynchronizationAL}.
We observed that the sequence $H_n$ also behaves in an almost linear periodic manner. Since $H_a$ is a private matrix, this behaviour does not compromise the protocol in an obvious manner. However, we also observed the sequence $M_n$ exhibit almost linear periodic behaviour. As $M_a$ is public, assuming that the $d$ and $\rho$ for $M_n$ are sufficiently small such that $M_{d+\rho}$ can be enumerated in polynomial time, it is possible to use this behaviour to derive $a$. $d$ and $\rho$ were sufficiently small in all combinations of $M$ and $H$ that we tested, as shown in Table \ref{tab:ExperimentResults}.

Below we describe an attack on the protocol that utilises the almost linear periodicity of $M_n$. The attack uses the public matrices $M$, $H$ and $M_a$ to derive the private exponent $a$. Once $a$ is obtained, it is trivial to use $M$, $H$, $M_b$, and $a$ with the procedures described in the protocol to obtain the secret key.
The attack is split into two parts; first, finding $d$ and $\rho$ for $M_n$, and second, finding $a$ using $d$ and $\rho$.

\subsection{Finding $d$ and $\rho$}
The sequence $M_n$ is defined by:
\begin{equation}
\begin{split}
M_1&=M\\
M_n&=(M_{n-1} \circ H) \oplus M \mbox{, for } n > 1\\
\end{split}
\end{equation}
$M$ and $H$ are public allowing for the enumeration of this sequence. Assuming $M_n$ is almost linear periodic, all terms after the defect that are a period apart differ by the linear factor. It follows that, after the defect, consecutive terms will change in a pattern that repeats every period. This implies that the sequence containing the differences between successive elements of $M_n$ will be periodic in nature after the defect.
We will refer to this sequence of differences between successive elements of $M_n$ as $D_n$:
\begin{equation}
    D_n=M_{n+1} - M_n,\mbox{ for }n\geq 1
\end{equation}
To find $d$ and $\rho$, we enumerate the terms of $D_n$ (by enumerating the terms of $M_n$ and calculating $D_n$) and compare the current term to previously enumerated terms. If a previously enumerated term is equal to the current term, it is possible that the previously enumerated term marks the beginning of periodicity, and its index is $d+1$. The difference in indexes of the previously enumerated term and the current term would, therefore, be equal to $\rho$. It is possible that terms could repeat before the defect, resulting in false values for $d$ and $\rho$. This is covered in more detail in \ref{specialcases}.
\subsection{Finding $a$}
\begin{enumerate}
\item From the definition of $D_n$:
\begin{equation}
    D_n=M_{n+1} - M_n,\mbox{ for }n\geq 1
\end{equation}
it follows that:
\begin{equation}
    M_n = M_1 + \sum_{i=1}^{n-1}D_i
\end{equation}
\item Let $Y=M_a-M_{d+1}$. It follows that:
\begin{equation}
\begin{split}
Y&=M_a-M_{d+1} \\
&=(M_1 + \sum_{i=1}^{a-1}D_i) - (M_1 + \sum_{i=1}^{d}D_i) \\
&=\sum_{i=d+1}^{a-1}D_i \\
\end{split}
\end{equation}
\item Due to the periodic nature of $D_i$, for $i>d$, this sum can be decomposed into two parts: the sum of the differences within the period, $\rho$, multiplied $x$ times, and the sum of the differences within the period in which $M_a$ is located up until $D_{a-1}$:
\begin{equation}
    Y=x\sum_{i=d+1}^{d+\rho}D_i + \sum_{i=a-k}^{a-1}D_i
\end{equation}
where $x$ is some positive integer, \(1\leq k\leq \rho\) and $d+x\rho+k=a$.

The periodic nature of $D_i$ after the defect implies that any sum of a number of consecutive elements in $D_i$, where the first term occurs after the defect, is equal to the sum of the same number of consecutive elements that occur any multiple of the period further along the sequence. Therefore the above can be rewritten as:
\begin{equation}
    Y=x\sum_{i=d+1}^{d+\rho}D_i + \sum_{i=d+1}^{d+k}D_i
\end{equation}
\item $k$ can be found by testing all possible values from $1$ to $\rho$. For a value to be $k$ the following must hold, for all indices $u,v$:
\begin{equation}
    (Y_{uv} - \sum_{i=d+1}^{d+k}D_{iuv})\mod\sum_{i=d+1}^{d+\rho}D_{iuv}=0
\end{equation}
\item Once $k$ has been obtained, $x$ can be found through the equation:
\begin{equation}
    \frac{Y_{uv} - \sum_{i=d+1}^{d+k}}D_{iuv}{\sum_{i=d}^{d+\rho}D_{iuv}}=x
\end{equation}
\item Now that we have $d$, $x$, and $k$, we can solve $d+x\rho+k=a$ to find $a$.
\end{enumerate}
\subsection{Special cases}
\label{specialcases}
There may be repeated elements in $D_n$, where $n\leq d$, leading to false values for $d$ and $\rho$. This will often be detected when searching for $k$, as no values for $k$ will satisfy the equation:
\begin{equation}
    (Y_{uv}-\sum_{i=a-k}^{a-1}D_{iuv})\mod\sum_{i=d}^{d+\rho}D_{iuv}=0
\end{equation}
There is a small chance that a value could satisfy the equation, resulting in an incorrect derivation of $a$. This can be handled by checking that the derived $a$ satisfies the equation:
\begin{equation}
    (M,H)^{derived\_a} = (M_a, Z)
\end{equation}
where the variable $Z$ can be ignored, because if the first term is correct, it follows that the second term is correct. If an incorrect $a$ is detected, the search for $d$ and $\rho$ can be resumed.

A second special case which should be accounted for in the attack, is when $D_n$ becomes the zero matrix for $n>d$. This results in a division by zero when finding $k$ and $x$. This special case is simple to account for, as it implies that for all $M_n$, such that $n>d$, $M_n=M_{d+1}$. Therefore, although it is impossible to find $a$, this has no bearing on the success of the attack, as $d+1$ can be substituted for $a$.

\subsection{Experimental Results}

\begin{table}[htb]
    \centering
    \begin{tabular}{|l|r|r|r|}
        \hline
        Maximum $d$         & 2151  \\
        Median $d$          & 26    \\
        Mean $d$            & 40.9  \\
        Maximal $\rho$      & 15    \\
        Median $\rho$       & 2     \\
        Mean $\rho$         & 2.8   \\
        Maximum attack time (s) & 200.9 \\
        Median attack time (s)  & 3.5   \\
        Mean attack time (s)    & 3.9   \\
        Success Rate        & 100\% \\
        \hline
    \end{tabular}
    \caption{Results of attack on the first protocol}
    \label{tab:ExperimentResults}
\end{table}

The success of the attack against $10000$ instances of the protocol, with the parameters suggested by \citeauthor{TropCrypt2}, is detailed in Table \ref{tab:ExperimentResults}. $d$ is the number of elements of the sequence $M^n$ that were enumerated before periodic behaviour was observed. $p$ is the period. The attack times give the time taken to find the private parameter $a$. The protocol and attack were implemented in Python and can be found in \cite{Implementation}. All tests were performed on a single core of an i7 CPU at 2.9GHz, with 8GB of RAM, running Windows 10, and interpreted using Python 3.7.6.
\section{Overview of the second protocol}
The protocol is based on a supposed semigroup (which we will show is not a semigroup in the next section) given by equipping the set of pairs, containing entries from $\overline{T}$, with the operation defined by:
\begin{equation}
    (M,G)(S,H)=((H\otimes M^T)\oplus (M^T\otimes H)\oplus S, G\otimes H)
\end{equation}
\begin{enumerate}
    \item Alice and Bob agree upon public matrices \(M\) and \(H\), with elements in $\mathbb{Z}$.
    \item Alice generates a private positive integer $a$, while Bob generates a private positive integer $b$.
    \item Alice calculates \((M,H)^a=(M_a,H_a)\), and shares $M_a$ with Bob, keeping $H_a$ private.
    \item Bob calculates \((M,H)^b=(M_b,H_b)\), and shares $M_b$ with Alice, keeping $H_b$ private.
    \item Alice calculates \(K=(M_b\otimes H_a)\oplus M_a\) to get the secret key.
    \item Bob calculates \(K=(M_a\otimes H_b)\oplus M_b\) to get the same key as Alice.
\end{enumerate}

As with the first protocol, \citeauthor{TropCrypt2} state that $(M,H)^a$ should be calculated using the square-and-multiply method.
\section{Proof that the second protocol cannot be implemented}
\label{sec:Analysis}
This protocol cannot be implemented as the operation that the protocol is based upon is not associative. Consider the example below:
\begin{equation}
\mbox{Let }A=
\begin{pmatrix}
0 & -1\\
0 & 0
\end{pmatrix},
B=
\begin{pmatrix}
0 & -2\\
0 & 0
\end{pmatrix},
\end{equation}

\begin{equation}
(A,B)^2=\left(
\begin{pmatrix}
-3 & -2\\
-1 & -3
\end{pmatrix},
\begin{pmatrix}
-2 & -2\\
0 & -2
\end{pmatrix}
\right)
\end{equation}

\begin{equation}
(A,B)(A,B)^2=\left(
\begin{pmatrix}
-3 & -2\\
-3 & -3
\end{pmatrix},
\begin{pmatrix}
-2 & -4\\
-2 & -2
\end{pmatrix}
\right)
\end{equation}

\begin{equation}
(A,B)^2(A,B)=\left(
\begin{pmatrix}
-4 & -5\\
-3 & -4
\end{pmatrix},
\begin{pmatrix}
-2 & -4\\
-2 & -2
\end{pmatrix}
\right)
\end{equation}

\begin{equation}
(A,B)(A,B)^2\neq (A,B)^2(A,B)
\end{equation}

It follows, from the operation's lack of associativity, that it is not possible to calculate \((M,H)^a\) by utilising the square-and-multiply method. Consequently, the protocol cannot be successfully implemented.
\section{Conclusion}
The first protocol we analysed is insecure when using the proposed parameters for key generation. It is unclear how to modify the protocol such that it resists the attack we describe. Our attack is significantly faster than the binary search attack given in \cite{RudyArxiv}, requiring about 0.5\% of the time to find $a$ when using proposed protocol parameters. The longest the attack took to break the protocol was $200$ seconds, which was still considerably faster than the binary search.

It is not possible to implement the second protocol we analysed, since the operation it relies upon is not associative. This prevents the use of the square-and-multiply method for exponentiation, which is a fundamental aspect of the general protocol on which these protocols are based \cite{kahrobaei2016using}.

We encourage interested readers to examine our implementation of tropical matrix algebra, the protocols, and the attack, and perform their own experiments using it \cite{Implementation}.
\printbibliography

@article{TropCrypt2,
  title={Tropical cryptography II: Extensions by homomorphisms},
  author={Grigoriev, Dima and Shpilrain, Vladimir},
  journal={Communications in Algebra},
  volume={47},
  number={10},
  pages={4224--4229},
  year={2019},
  publisher={Taylor \& Francis}
}

@article{Kotov2015AnalysisOA,
  title={Analysis of a key exchange protocol based on tropical matrix algebra},
  author={Matvei Kotov and Alexander Ushakov},
  journal={Journal of Mathematical Cryptology},
  year={2015},
  volume={12},
  pages={137 - 141}
}

@misc{Implementation,
  author = {Isaac, Steve},
  title = {Implementation of protocol and attack},
  howpublished = {\\\url{https://github.com/steveisaac/TropicalCryptography}},
}

@InProceedings{SemidirectProductofSemigroups,
author="Habeeb, Maggie
and Kahrobaei, Delaram
and Koupparis, Charalambos
and Shpilrain, Vladimir",
title="Public Key Exchange Using Semidirect Product of (Semi)Groups",
booktitle="Applied Cryptography and Network Security",
year="2013",
publisher="Springer Berlin Heidelberg",
pages="475--486",
isbn="978-3-642-38980-1"
}

@inproceedings{kahrobaei2016using,
  title={Using semidirect product of (semi) groups in public key cryptography},
  author={Kahrobaei, Delaram and Shpilrain, Vladimir},
  booktitle={Conference on Computability in Europe},
  pages={132--141},
  year={2016},
  organization={Springer}
}

@book{Max-linearSystems,
  title={Max-linear systems: theory and algorithms},
  author={Butkovi{\v{c}}, Peter},
  year={2010},
  publisher={Springer Science \& Business Media}
}

@article{MatrixPowers,
  title={Powers of matrices over an extremal algebra with applications to periodic graphs},
  author={Nachtigall, Karl and others},
  journal={Mathematical Methods of Operations Research},
  volume={46},
  number={1},
  pages={87--102},
  year={1997},
  publisher={Springer \& Gesellschaft f{\"u}r Operations Research (GOR) \& Nederlands~…}
}

@article{Baccelli1994SynchronizationAL,
  title={Synchronization and Linearity: An algebra for discrete event systems},
  author={François Baccelli and Gregg A. Cohen and G. J. Olsder and Jean-Pierre Quadrat},
  journal={Journal of the Operational Research Society},
  year={1994},
  volume={45},
  pages={118-119}
}

@ARTICLE{RudyArxiv,
       author = {{Rudy}, Dylan and {Monico}, Chris},
        title = "{Remarks on a Tropical Key Exchange System}",
      journal = {arXiv e-prints},
         year = 2020,
        month = may,
archivePrefix = {arXiv},
       eprint = {2005.04363},
 primaryClass = {cs.CR},
       adsurl = {https://ui.adsabs.harvard.edu/abs/2020arXiv200504363R}
}
\end{document}